# Assessing Gender Bias in Predictive Algorithms using eXplainable AI*

Cristina Manresa-Yee and Silvia Ramis
Maths and Computer Science Department
Universitat de les Illes Balears
Palma, Spain
{cristina.manresa, silvia.ramis}@uib.es

## ABSTRACT

Predictive algorithms have a powerful potential to offer benefits in areas as varied as medicine or education. However, these algorithms and the data they use are built by humans, consequently, they can inherit the bias and prejudices present in humans. The outcomes can systematically repeat errors that create unfair results, which can even lead to situations of discrimination (e.g. gender, social or racial).

In order to illustrate how important is to count with a diverse training dataset to avoid bias, we manipulate a well-known facial expression recognition dataset to explore gender bias and discuss its implications.

## CCS CONCEPTS

• Human-centered computing ~ Human computer interaction (HCI)

## 1 Introduction

In 2015, Amazon realized that their algorithm used for recruiting employees was biased against women. Amazon's CV screening tool was trained on biased historical data of submitted CVs, which reflected the male dominance across the tech industry [7].

With the widespread of Artificial intelligence (AI) in real world applications such as healthcare or finances, the need to understand the rationale or examine the internal working of the system is fundamental. Indeed, the EU directive 2016/680 General Data Protection Regulation (GDPR) highlights the importance of the Right to Explanation, that is, "a right that individuals might exercise when their legal status is affected by a solely automated decision" [23]. Applied to AI, individuals have the right to be given an explanation for an output of the algorithm. Explanations can be given before or after decisions are made, and the kind of explanations will inform about the dataset used in the training or will be focused on understanding the model behavior or a specific instance prediction.

Predictive systems are based on statistical tools, AI algorithms and machine learning models to make predictions or decisions. We expect automated decision-making systems to be fair, neutral and impartial [6]. However, these systems depend on factors such as training datasets or a model design. Therefore, any unintentionally inherit latent biases included will perpetuate in the results [14] and can reflect societal asymmetries, such as racial, social class or gender bias. These biases can be a problem when they discriminate systematically certain groups of people.

To study and analyze the explaining dimension of AI, we find the research field of eXplainable AI (XAI). XAI can be defined as "AI systems that can explain their rationale to a human user, characterize their strengths and weaknesses, and convey an understanding of how they will behave in the future" [12]. This is especially valuable when the model underneath the AI system is considered a black-box (e.g. deep learning).

To discuss the importance of the dataset in gender bias, we use a case study related to face emotion recognition, which is broadly applied in many fields such as robotics, medicine or marketing [26]. The aim of this work is to build a Convolutional Neural Network (CNN) to classify Ekman's six basic emotions (angry, fear, disgust, sadness, surprise and happiness) [9] to examine the importance of a balanced dataset and explore the similarities and differences between emotions posed by female or male individuals. We explore diverse combinations of training and testing datasets to assess gender bias in datasets and highlight the importance of counting with diversity in the training sets.

The work is organized as follows: Section 2 describes the XAI research line and compiles works of intelligent systems related to gender bias. Section 3 presents the model used to classify, the pre-processing steps, the dataset and the XAI approach. Section 4 describes the experiments, and the results and discussion are commented in Sections 5 and 6 respectively. Finally, we summarize the findings and contributions from the evaluation and implications for future work.

## 2 Related work

In this section, we present XAI characteristics from a HCI perspective and research works that have studied gender bias before.

### 2.1 eXplainable Artificial Intelligence (XAI)

Although interpreting intelligent systems has a long history, recently, researchers in HCI seek a more human-centered approach: analyzing how humans explain to each other to replicate it for AI systems [18], informing design practices and developing frameworks for XAI [2, 16, 27], or evaluating explanations with humans in XAI [4].

Explainability in predictive systems encompasses concepts such as fairness, causality, transparency, reliability or trust [15]. Therefore, one of the motivations for XAI is related with this work, that is, identifying bias within training data, models or deployed system [3].

Explanations should be delivered depending on the end user (e.g. e.g. domain experts, users affected by the outcome of the system, regulatory entities or developers) [2, 12]) and therefore different methods exist to improve the explainability of the models and to design explanation interfaces [25].

### 2.2 Gender Bias

Although bias is associated with a negative connotation, bias is just a deviation from the standard. Danks and London [6] identified five algorithmic biases in automated systems:

1. Training data bias: automated systems rely on training data, therefore, if the input data is biased in one or another way, the outcome can also reflect the bias.
2. Algorithmic focus bias: this bias is included through differential usage of information in the input or training data, like for example the deliberately non-use of certain information, even if it is available.
3. Algorithmic processing bias: this bias arises when the algorithm itself is somehow biased. An example is the use of a statistically biased estimator.
4. Transfer context bias: this bias is related to the use of an algorithm in a context for which it was not modelled.
5. Interpretation bias: this bias is the misinterpretation of the algorithm's outputs or functioning by the user.

Any occurrence of these biases when systematically affects negatively to a group of people, may lead to discrimination – which is legally defined "as the unfair or unequal treatment of an individual (or group) based on certain characteristics such as income, education, gender or ethnicity" [1].

We find multiple evidence of gender bias in AI systems in different domains, that exhibit gender disparities.

Regarding language translation, Prates et al. [19] showed that statistical translation tools such as Google Translate can exhibit gender biases and a strong tendency toward male defaults. They translated professional-related sentences such as "He/She is an engineer" and adjective-related sentences including adjectives such as Happy or Shy from 12 gender neutral languages (e.g. Malay or Hungarian) to English. Then, they analyzed the statistics between female and male pronominal genders in the translations. They provided evidence that male defaults were prominent, especially in fields such as STEM (Science, Technology, Engineering and Mathematics) occupations. Regarding the adjectives, there was also a bias. Adjectives such as Shy and Desirable were linked with a larger proportion to female pronouns, whereas Guilty and Cruel, were translated almost exclusively with male ones. In the case of Google Translate, a newest version made efforts to reduce this bias, therefore, currently, they present both the feminine and masculine translations.

Lambrecht and Tucker [13] analyzed how delivered ads promoting job opportunities in STEM fields were shown more often to men than to women. The ad was explicitly intended to be gender-neutral in its delivery, however, the algorithm was modeled to optimize cost-effectiveness, and consequently, the ad was shown more to men, as young women are a prized target and more expensive to show ads to.

In the healthcare domain, specifically in Precision Medicine, Cirillo et al. [5] observed that AI could sometimes neglect desired differentiations such as sex and gender. In this domain, XAI could help to justify clinical predictions and decisions when they are differential for patients with different sex and genders (e.g. lack of balanced sex and gender representation data, understanding the differences that are representative to promote desired bias). Examples of including or removing the sex and gender information could impact for example in diagnosing coronary artery illness is women.

Similar to the work we present, Domnich and Anbarjafari [8] investigated how gender bias can affect face emotion recognition. They experimented with different model architectures and training-testing datasets to analyze the fairness in the datasets. They evaluated which were the most biased architectures regarding fairness and what kind of emotions were easier to recognize for men or women. To simplify the classification, they worked with four facial expressions: happiness, sadness, surprise and upset (that united contempt, disgust, and anger to relax the difficulty). In our case, we just use one model to manipulate the datasets and study the outcomes, but we include Ekman's six basic emotions: (angry, fear, disgust, sadness, surprise and happiness) [9].

# 3 Methodology

We built a CNN model in Keras to classify the six basic emotions identified by Ekman. CNN are considered a black-box model [12].

## 3.1 Data description

We describe the BU-4DFE Dataset, a well-known public dataset and widely used as benchmark in Facial Expression Recognition. The BU-4DFE Dataset contains 101 subjects, 58 females and 43 males with a variety of ethnic ancestries (asian, black, hispanic, and white). The subjects are non-professional actors and for each subject, there are six sequences of video showing the six facial expressions (anger, disgust, fear, happiness, sadness, and surprise), respectively. Each sequence is about a minute long. Teixera Lopes et al. [24] studied different datasets for facial expression recognition. They obtained an accuracy of 72.89% with BU-4DFE, and although they reported higher accuracies for other datasets like CK+, these are not labelled with gender information. In the case of BU-4DFE, all files' names include the information on the gender of the subject.

## 3.2 Data pre-processing

The pre-processing step includes data homogenization and augmentation. First the face is detected using the method proposed in [17]. Then, we align the images to eliminate rotations and achieve uniformity between them. Finally, the face is cropped, converted to grayscale in range from 0 to 255 and resized to the size of the input data that the CNN needs.

On the other hand, the CNN need to have a large enough number of training samples that must contain significant facial variations. Therefore, we perform an augmentation step where we add small variations in terms of lighting and appearance. We use the gamma correction technique to vary the illumination [11], selecting three values that continue allowing the clearly distinguishment of the face. In this way, we quadruplicate the data (three gamma values plus the original one). Another variation introduced to augment data is to translate four pixels in both axis and crop the image (where the face is always present with two eyes, nose and mouth). Finally, we duplicate the images through a horizontal flip.

## 3.3 Model

We build a fine-tuned CNN model [20], which is depicted in Fig. 1. This CNN receives as input a 150x150 image and classifies it into one of the six emotions. The architecture consists of 5 convolutional layers, 3 pooling layers and two fully connected layers. The first fully connected layer also has a dropout [10] to avoid overfitting in the training.

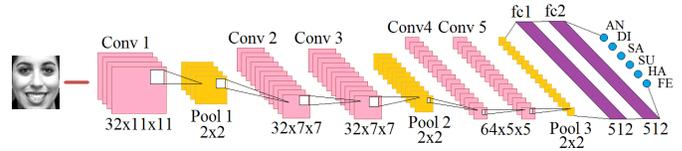

**Figure 1. Architecture of the CNN to classify six emotions: Anger (AN), Disgust (DI), SA (Sadness), SU (Surprise), HA (Happiness) and FE (Fear)** [20]

## 3.4 XAI approach

To analyze visually the outcome of the system, we use LIME, Local Interpretable Model-agnostic Explanation [21]. LIME offers locally faithful explanations within the surroundings instance being explained; therefore, it does not explain the global behavior of the model but explains how the prediction was made. One of the advantages of LIME is that it can be used with all classifiers and can process tabular data, text, or images. In the case of images, it shows those regions, superpixels, that have more importance in the classification.

# 4 Experiment

The experiment section includes details on the training and testing datasets as well as the followed procedure.

## 4.1 Training

For training the CNN, we take the 80% of the data included in the dataset.

BU-4DFE dataset contains 58 female and 43 male subjects. For each subject, there is a video of approximately one minute posing each of the emotions, where generally, all videos start with the individual posing the neutral facial expression, and from the second 20 they change the expression to the one requested.

Therefore, to create the instances for the test group, we take snapshots between second 20 and second 40 with five seconds of interval, so that there are small variations of the same expression. We then review the images carefully to verify the facial expression. If the actor or actress has not yet performed the facial expression in the second 20 and poses still the neutral face, the image is removed. Thus, for each expression of each subject there are about five images, and the pre-processing step is applied to each image.

In order to analyse the effect of bias on recognition, once applied the pre-processing step, we use the training dataset in three ways: (1) In the first training dataset with about 23000 images, each class contains males and females. (2) In the second training dataset with about 8000 images, each class contains only males. (3) And the third training dataset with about 15000 images, each class contains only females.

## 4.2 Testing

For the testing, we take the 20% of the data. It is important to highlight, that these data are not included in the training dataset.

The BU-4DFE testing dataset contains an average of 18 images for each expression of each subject and we apply the pre-processing step to each image (without the steps for data augmentation): detect the face, align it, crop it, convert it to grayscale, and resize it to the size of the input data required by the CNN.

Once applied the pre-processing step, we prepare three testing datasets with six classes (anger, disgust, fear, happiness, sadness and surprise): (1) First testing dataset with 107 images, each class contains males and females. (2) Second testing dataset with 49 images, each class contains only males. (3) And the third training dataset with 58 images, each class contains only females.

### 4.3   Procedure

With the data grouped as described in the aforementioned subsections, we train three neural networks to classify the six basic facial expressions:

- The first CNN is trained with the training dataset that contained images of both females and males for each facial expression.
- The second CNN is trained with the training dataset that contained images of only males for each facial expression.
- The third CNN is trained with the training dataset that contained images of only females for each facial expression.

The model is built on Keras, an open-source deep learning library, and all trainings are performed on a GPU is NVIDIA Tesla-K40C with 12 GB of memory. For the training of the CNN, we have taken into account the size of the input data of the neural network. We highlight that the aim of the study is to explore gender bias, without focusing on improving the accuracy of the model.

We test the datasets to obtain the results grouping and combining the training and testing datasets with males, females or both (See Table 1). Finally, we apply LIME to observe the face regions that are important for the model to classify images into an emotion class. In this case, LIME is configured to show the 10 most important features for the classification.

**Table 1: Summary of experiments**

| Training groups (80%) | Testing groups (20%) |
|---|---|
| Both female and male (B) | Both females and males |
|  | Only females |
|  | Only males |
| Only females (F) | Only females |
|  | Only males |
| Only males (M) | Only females |
|  | Only males |

## 5   Results

Results are divided in two subsections. First, we analyze the confusion matrices, and then, we assess the zones that the CNN consider more relevant for its decisions.

### 5.1   Confusion matrices

The model trained both with females and males obtain decent performance on all testing datasets, being Surprise, Angry and Happiness the emotions better classified. In all cases, the model shows excellent recognition for all testing datasets for the Surprise emotion and for Happiness also in the case of females. In the case of males, the system misclassifies frequently Sadness with Angry and the Fear emotion is widely confused with other emotions such as Angry, Happiness or Sadness. However, regarding females, Sadness is confused both with Angry and Fear. And surprisingly, Fear is very confused with Happiness (see Fig. 2).

When training with male facial expressions, Surprise and Sadness are well classified by the model both for men and women. The Sad images in the male training dataset seem to be more informative, as both female and male testing datasets improve their accuracy in regard to all the experiments. However, the accuracy for the Fear emotion in women is much lower than in men, as we observe a misclassification with Happy. There is a significant decrease in the recognition of the Anger emotion regarding Fig. 2 (See Fig. 3) both for female and male. In the case of females, it highly misclassifies it with Sadness, and in the case of males, with Disgust and also with Sadness. Similarly, the Happy expression for female decreases in accuracy, and it is confused with Fear. On the contrary, Sadness is classified with higher accuracy than when trained with men and female.

| B-B | Angry | Disgust | Fear | Happy | Sad | Surprise |
|---|---|---|---|---|---|---|
| Angry | 80.95 | 4.76 | 0.00 | 4.76 | 9.52 | 0.00 |
| Disgust | 4.76 | 57.14 | 9.52 | 28.57 | 0.00 | 0.00 |
| Fear | 12.50 | 0.00 | 12.50 | 62.50 | 12.50 | 0.00 |
| Happy | 0.00 | 10.53 | 10.53 | 73.68 | 5.26 | 0.00 |
| Sad | 40.00 | 0.00 | 6.67 | 0.00 | 53.33 | 0.00 |
| Surprise | 0.00 | 0.00 | 0.00 | 0.00 | 0.00 | 100.00 |

| B-M | Angry | Disgust | Fear | Happy | Sad | Surprise |
|---|---|---|---|---|---|---|
| Angry | 81.82 | 0.00 | 0.00 | 0.00 | 18.18 | 0.00 |
| Disgust | 11.11 | 66.67 | 22.22 | 0.00 | 0.00 | 0.00 |
| Fear | 28.57 | 0.00 | 14.29 | 28.57 | 28.57 | 0.00 |
| Happy | 0.00 | 20.00 | 20.00 | 50.00 | 10.00 | 0.00 |
| Sad | 42.86 | 0.00 | 0.00 | 0.00 | 57.14 | 0.00 |
| Surprise | 0.00 | 0.00 | 0.00 | 0.00 | 0.00 | 100.00 |

| B-F | Angry | Disgust | Fear | Happy | Sad | Surprise |
|---|---|---|---|---|---|---|
| Angry | 80.00 | 10.00 | 0.00 | 10.00 | 0.00 | 0.00 |
| Disgust | 0.00 | 50.00 | 0.00 | 50.00 | 0.00 | 0.00 |
| Fear | 0.00 | 0.00 | 11.11 | 88.89 | 0.00 | 0.00 |
| Happy | 0.00 | 0.00 | 0.00 | 100.00 | 0.00 | 0.00 |
| Sad | 37.50 | 0.00 | 12.50 | 0.00 | 50.0 | 0.00 |
| Surprise | 0.00 | 0.00 | 0.00 | 0.00 | 0.00 | 100.00 |

**Figure 2. Test with the training that contains female and male. (B-B) the test dataset contains both genders. (B-M) the test dataset contains only males. (B-F) the test dataset contains only females.**

| M-M | Angry | Disgust | Fear | Happy | Sad | Surprise |
|---|---|---|---|---|---|---|
| Angry | 27.27 | 45.45 | 0.00 | 0.00 | 27.27 | 0.00 |
| Disgust | 0.00 | 55.56 | 22.22 | 11.11 | 11.11 | 0.00 |
| Fear | 0.00 | 0.00 | 71.43 | 28.57 | 0.00 | 0.00 |
| Happy | 0.00 | 10.00 | 40.00 | 50.00 | 0.00 | 0.00 |
| Sad | 0.00 | 0.00 | 14.29 | 0.00 | 85.71 | 0.00 |
| Surprise | 0.00 | 0.00 | 0.00 | 0.00 | 0.00 | 100.00 |

| M-F | Angry | Disgust | Fear | Happy | Sad | Surprise |
|---|---|---|---|---|---|---|
| Angry | 10.00 | 0.00 | 10.00 | 10.00 | 70.00 | 0.00 |
| Disgust | 0.00 | 50.00 | 16.67 | 25.00 | 0.00 | 8.33 |
| Fear | 0.00 | 0.00 | 33.33 | 66.67 | 0.00 | 0.00 |
| Happy | 0.00 | 0.00 | 33.33 | 66.67 | 0.00 | 0.00 |
| Sad | 0.00 | 0.00 | 12.50 | 0.00 | 87.50 | 0.00 |
| Surprise | 0.00 | 0.00 | 0.00 | 0.00 | 0.00 | 100.00 |

**Figure 3. Test with the training that contains only males. (M-M) the test dataset contains only males. (M-F) the test dataset contains only females**.

The training with females shows excellent classification for female faces regarding Happiness and Surprise, similar as in Fig. 2, which can imply that there is a relevant number of images for both expressions in the case of females due to the clarity of these expressions. The Fear expression for females is also low using this training, misclassifying it with Disgust and Happy (see Fig. 4). In all cases, we have obtained a very low rate of correct classifications for the Fear emotion in female, this can imply that there is a lack of images for this emotion due to its great variability in posing it and the difficulty of posing a real Fear emotion by non-professional actors (see Fig. 5).

| F-F | Angry | Disgust | Fear | Happy | Sad | Surprise |
|---|---|---|---|---|---|---|
| Angry | 50.00 | 20.00 | 10.00 | 0.00 | 0.00 | 20.00 |
| Disgust | 0.00 | 50.00 | 0.00 | 25.00 | 0.00 | 25.00 |
| Fear | 0.00 | 22.22 | 11.11 | 66.67 | 0.00 | 0.00 |
| Happy | 0.00 | 0.00 | 0.00 | 100.00 | 0.00 | 0.00 |
| Sad | 25.00 | 0.00 | 12.50 | 0.00 | 62.5 | 0.00 |
| Surprise | 0.00 | 0.00 | 0.00 | 0.00 | 0.00 | 100.00 |

| F-M | Angry | Disgust | Fear | Happy | Sad | Surprise |
|---|---|---|---|---|---|---|
| Angry | 90.91 | 0.00 | 0.00 | 0.00 | 9.09 | 0.00 |
| Disgust | 0.00 | 77.78 | 11.11 | 0.00 | 11.11 | 0.00 |
| Fear | 28.57 | 28.57 | 42.86 | 0.00 | 0.00 | 0.00 |
| Happy | 10.00 | 40.00 | 0.00 | 30.00 | 20.00 | 0.00 |
| Sad | 42.86 | 0.00 | 0.00 | 0.00 | 57.14 | 0.00 |
| Surprise | 0.00 | 0.00 | 20.00 | 0.00 | 0.00 | 80.00 |

**Figure 4. Test with the training that contains only females. (F-F) the test dataset contains only females. (F-M) the test dataset contains only males.**

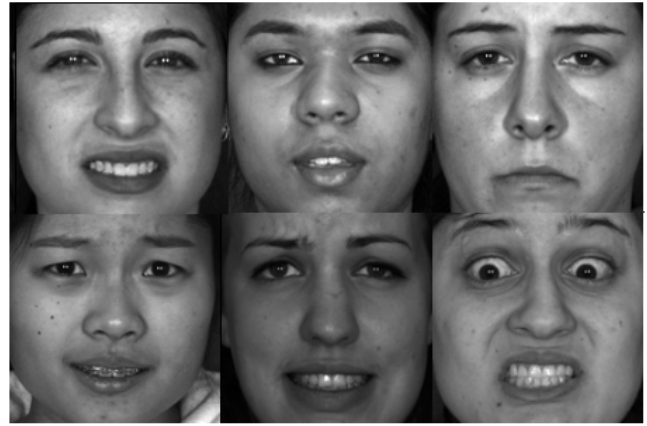

**Figure 5. Excerpt of images labelled as Fear from BU-4DFE**

## 5.2 LIME

Analyzing the confusion matrices, we use LIME to explore more exhaustively the misclassifications. First, in Fig. 6, we show examples of the important features used by the CNN to correctly classify each emotion.

A global observation in the experiments is that Fear is frequently misclassified with Happiness, especially in the case of female testing datasets. The reason may be that in the misclassified images, subjects show their teeth, and the outer corners of the lips may be slightly raised (like the images classified as Happiness). The Fear emotion may pose similar characteristics, especially when it is performed by a non-professional actor. Fig. 7 shows different images not correctly classified. The regions considered by the CNN are the mouth (with appearance of teeth), the cheeks raised (framed with the expression lines) and the brows-forehead zone.

Disgust shares characteristics with other expressions such as Fear and Angry (brows drawn or pulled down). In Fig. 8, we show three different expressions misclassified as Disgust, due to the common regions. From a human perspective, the Happy expression misclassified, would also be considered a non-happy expression, therefore, is sensible that the model classified it with other emotion. Similar reasoning could also be applied to the Fear and Angry expressions, both could also be misclassified by a human. In all three images, LIME is highlighting the brows-forehead zone and a slightly open or closed mouth zone.

Sadness and Anger are emotions that are confused between them. The visualizations generated with LIME show that the region between the eyes is important in both emotions (see Fig.9).

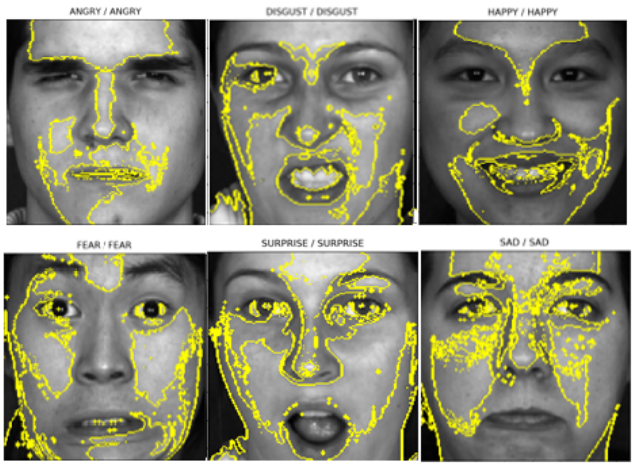

**Figure 6. LIME applied on correctly labelled images**

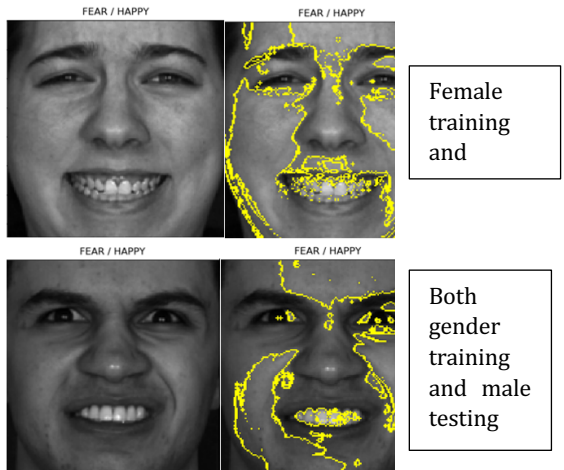

**Figure 7. Misclassified images with happiness .**
**Figure 8. Misclassified images with disgust**

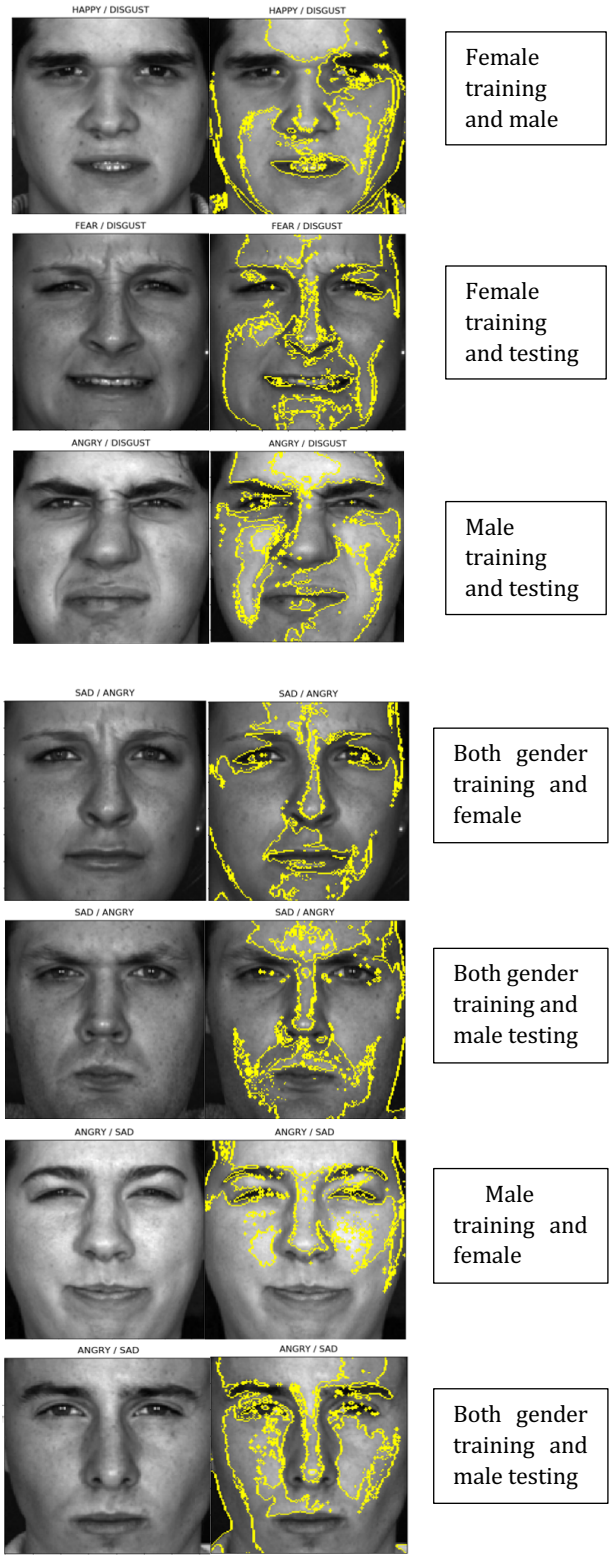

**Figure 9. Misclassified images Anger-Sadness (and opposite)**

**Table 2: Misclassified emotions. Training only with females**

|  | Test with only Females (Misclassified facial expressions) | Test with only Males (Misclassified facial expressions) |
|---|---|---|
| Angry | Disgust, Fear, Surprise | Sad |
| Disgust | Happy, Surprise | Fear, Sad |
| Fear | Disgust, Happy | Disgust, Angry |
| Happy |  | Angry, Disgust, Sad |
| Surprise |  | Fear |

**Table 3: Misclassified emotions. Training only with males**

|  | Test with only Females (Misclassified facial expressions) | Test with only Males (Misclassified facial expressions) |
|---|---|---|
| Angry | Fear, Happy, Sad | Disgust, Sad |
| Fear | Happy | Happy |

## 6 Discussion

There are some emotions that are difficult to distinguish between them such as disgust and anger, even for humans [22]. Specially in the BU-4DFE, the fear emotion in females seems difficult to classify. Regarding the misclassifications between females and males, it is surprising that for a particular emotion, the misclassifications with other emotions are different in females or males (see Table 2 and 3).

Analyzing the overall accuracies, in the case of Happiness, all training favored the recognition in females, while in males, this emotion was misclassified with others, especially when the training dataset only contained female.

When training with only males, we can also appreciate differences between males and females. Although both genders share common misclassification in some emotions, such as Anger with Sadness and Fear with Happy, in the case of females this misclassification is higher than in males. The 70% of the female images labeled with Anger were classified as Sadness, whilst the male images only misclassified 27.27%, and the same happens with the misclassification of Fear with Happy. The 66.67% of the female images labeled with Fear were classified as Happiness, whilst the male images only misclassified 28.58%.

The relevant features of the images shown by LIME help understanding the misclassifications of emotions. However, analyzing the images, the model's outcomes seem sensible, and humans may also perceive different emotions as the ones labelled in the dataset. Therefore, the use of professional actors in the creation of a dataset or selecting images in the wild that pose real emotions is fundamental to achieve a high-quality dataset.

Observing in detail the important regions used by the CNN between males and females it can also inform of differences in the training dataset. Just as example, Fig. 10 show the regions for Anger in correctly labelled images trained with males and females separately: male training seems to give more importance than female to the brows-forehead region, and female training also considers the lower face region. This explains some remarkable misclassifications, such as Happy with Anger (see Fig. 11) in the case of the male testing with the female training. In Fig. 11, we depict the original image and the important regions for the CNN to classify it into one emotion. The regions marked on the male image labelled as Happiness (see Fig. 11), are similar to those marked to classify as Anger achieved with the female training (see Fig. 10). Therefore, the misclassification done by the CNN seems to be reasonable.

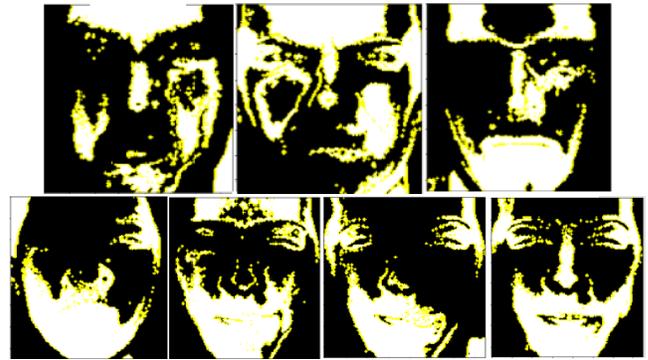

**Figure 10. Anger images correctly classified. Top row: male training. Bottom row: Female training**

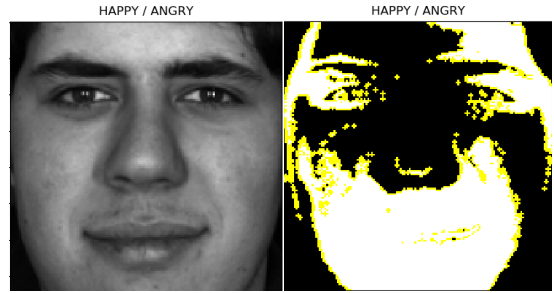

**Figure 11. Male image labelled with Happiness misclassified with Anger. This image has been tested with the Female training.**

## 7 Conclusions

In this work, we explored the impact of gender bias in training datasets for predictive algorithms. We used as case study a system that recognizes Ekman's six basic emotions: angry, fear, disgust, sadness, surprise, and happiness and worked with the well-known and popular facial expression recognition dataset BU-4DFE. We manipulated the training dataset to include both genders, only females or only males, and tested the model with new images also grouped by gender. To understand the misclassifications, we applied LIME, a XAI method that help us to make particular predictions comprehensible.

Although the facial expressions count with universal characteristics (even among ethnicities), we found some differences regarding misclassifications, recognitions rates or important regions that the model uses for classifying when training with males or females. The dissimilarities in the results acknowledge the importance of counting with a high-quality training dataset with diversity to reduce biases. In particular, the BU-4DFE dataset lack of more samples for some difficult emotions such as Fear or Disgust, that can be posed with a high variety of facial expressions (specially by non-professional actors).

The number of intelligent systems in everyday life is increasing rapidly and they are getting more sophisticated. Besides accuracy rates, people need further explanations to understand the decisions and predictions made by the system. XAI opens a door to offer these explanations, which contribute to achieve transparent and responsible AI and helps people to identify biases, trust the system and make informed decisions.

Identifying and removing undesirable bias from predictive algorithms (starting from the data used to train the model) can be a powerful way to avoid gender inequalities.

Future studies could investigate the biases (gender or ethnical) present in other popular facial expression recognition datasets, and we could also transfer this methodology to other domains.

## ACKNOWLEDGMENTS

This work has been supported by the Agencia Estatal de Investigación, project PID2019-104829RA-I00 / AEI / 10.13039/501100011033, EXPLainable Artificial INtelligence systems for health and well-beING (EXPLAINING).